\documentclass{sig-alternate}

\newcommand {\ra} {\rangle}
\newcommand {\la} {\langle}

\renewcommand{\thesection}{\arabic{section}}

\newcommand{\ket}[1]{\left|#1\right>}

\newtheorem{lemm}{Lemma}[section]
\newtheorem{lemma}[lemm]{Lemma}

\newtheorem{coro}[lemm]{Corollary}

\newtheorem{deff}[lemm]{Definition}
\newtheorem{defn}[lemm]{Definition}

\newtheorem{thm}[lemm]{Theorem}
\newtheorem{theo}[lemm]{Theorem}
\newtheorem{claim}[lemm]{Claim}

\newcommand{\C}{{\cal C}}

\newcommand{\beq}[1]{\begin{equation}\label{#1}}
\newcommand{\enq}[0]{\end{equation}}

\newcommand{\Nn}[0]{{\bf N}}

\newcommand{\h}[0]{{\cal H}}

\newcommand{\ignore}[1]{}

\begin{document}

\conferenceinfo{STOC'01,} {July 6-8, 2001, Hersonissos, Crete, Greece.}
\CopyrightYear{2001}
\crdata{1-58113-349-9/01/0007}

\title{Quantum Walks on Graphs}
\author{Dorit Aharonov\thanks{E-mail: doria@cs.berkeley.edu, 
 Computer Science Division, U.C. Berkeley, Berkeley, California, 
USA, supported by U.C. President's postdoctoral fellowship and
 NSF grant CCR-9800024}\hspace{6mm}
 Andris Ambainis\thanks{E-mail: ambainis@cs.berkeley.edu, 
 Computer Science Division, U.C. Berkeley, Berkeley, California, USA,supported by Microsoft Graduate Fellowship and NSF grant CCR-9800024}\hspace{6mm} 
 Julia Kempe\thanks{E-mail: kempe@math.berkeley.edu, Departments of Mathematics and Chemistry, U.C. 
Berkeley, Berkeley, California, USA, supported by the Center for Pure
and Applied Mathematics, U.C. Berkeley, and by NSA and ARDA under ARO}\hspace{6mm}
 Umesh Vazirani \thanks{E-mail: vazirani@cs.berkeley.edu, 
 Computer Science Division, U.C. Berkeley, Berkeley, California, USA,
supported by NSF grant CCR-9800024}}

\maketitle
\begin{abstract}
We set the ground for a theory of quantum walks on graphs- 
the generalization of random walks on finite
graphs to the quantum world. Such quantum walks do not converge 
to any stationary distribution, as they are unitary and reversible.  
 However, by suitably relaxing the definition, we
can obtain a measure of how fast the quantum walk spreads or how confined
the quantum walk stays in a small neighborhood. We give definitions of
mixing time, filling time, dispersion time.
We show that in all these measures, the quantum walk on the cycle 
is almost quadratically faster then its classical correspondent.
On the other hand, we give a lower bound on the possible speed up 
by quantum walks for general graphs, showing that
 quantum walks can be at most polynomially faster than their classical 
counterparts. 
\end{abstract}
\section{Introduction}
Markov chains or random walks on graphs have proved to be a fundamental
tool, with broad applications in various fields of mathematics, 
computer science and the natural sciences, 
such as mathematical modeling of physical systems,
simulated annealing, and the Markov Chain 
Monte Carlo method. In the physical sciences they provide a fundamental
model for the emergence of global properties from local interactions.
In the algorithmic context, they provide a general paradigm for sampling 
and exploring an exponentially large set of combinatorial structures
(such as matchings in a graph), by using a sequence of simple, local
transitions.

In this paper, we initiate a study of the theory of quantum walks on graphs ---
the motivation, as in the case of Markov chains, is to study 
global properties of a certain structured set,
using repeated application of local transition
rules. In the quantum setting, though, 
the local transition rule is defined to be unitary, rather than 
probabilistic. A classical Markov chain is said to be 
a random walk on 
an underlying graph, if the nodes of the graph are the states in $S$, 
and a state $s$ has non zero probability to go to $t$
if and only if the edge $(s,t)$ exists in the graph. To define a 
quantum random walk, in addition to the Hilbert space spanned by the nodes
of the graph, we must explicitly introduce the Hilbert space spanned by
the outcomes of the coin that control the process.  Thus,  
the quantum walk is allowed to use an auxiliary Hilbert space.
Now, the quantum
walk on a graph is naturally defined to be a unitary transformation
on the tensor product of the Hilbert space of the graph and the auxiliary 
Hilbert space, and with the property
that the probability amplitude (rather than the
probability) is non zero only on edges of the graph.

How do the basic definitions of Markov chains 
carry over to quantum walks? 
The most fundamental property of Markov chains is
the fact that they converge to
a stationary distribution, independent of the initial state. 
However, by their very definition, quantum walks do not converge 
to any stationary state.  This is due to the fact that
unitary matrices preserve the norm 
of vectors, and hence the distance between the vectors describing 
the system at subsequent times  does not converge to $0$. 
One can ask whether the probability 
distribution induced on the nodes of the graph converges in time, 
but it turns out that it does not converge either. 
Yet we can obtain a natural notion of convergence   
in the quantum case, if we define 
the limiting distribution as the limit of the {\it average} of the probability 
distributions over time. This definition captures  
the amount of time the walk spends in each subset of the nodes,
 and moreover, it corresponds to the natural concept
 of sampling from the graph, since if one measures the state at a random 
time chosen from the interval $\{1,..,t\}$, 
  the resulting distribution is exactly the average distribution.
We show that although
in general, the limiting distribution is a function of the initial state
of the quantum walk, for Cayley graphs of Abelian groups
 it is independent of the initial state, 
and is uniform over the group elements.

The rate of convergence, called the mixing time, is of crucial
importance to algorithmic applications of classical Markov chains.
Given the notion of limiting distribution in the quantum case, 
we can now talk about mixing times 
of a quantum walk. A natural definition for mixing time 
is the time it takes 
for the average probability distribution to get close to 
the limiting distribution. 
We can also talk about measures for
 how fast the quantum walk spreads or how long it takes the quantum
 walk to escape from a small neighborhood. 
We give definitions of quantum mixing time, sampling time, 
filling time, and dispersion time. 
How do the various mixing times of quantum walks compare with their classical
counterparts? We show that the quantum walk on a cycle converges in time
$O(n \log n)$, giving a nearly quadratic speedup over the classical
walk. For the cycle this quadratic speed up is the best possible, 
since the diameter of the graph is clearly a lower bound for 
the mixing time. How large can the quantum speed up be, for other 
graphs? We give a general lower bound on the various measures for the 
quantum mixing time, in terms of the conductance of the underlying graph. 
Our main result is that quantum random walks on graphs can be at most 
polynomially faster than their classical counterparts, 
and in fact, for bounded degree graphs, 
the gap is at most quadratic.

It is still an open question whether quantum walks can be used to 
obtain a quadratic speed up for certain randomized algorithms --- such as 
2-SAT. Indeed, all quantum algorithms from the last decade ---
including Shor's celebrated factorization algorithm\cite{shor}
 and Grover's search
algorithm\cite{grover} ---
 use only quantum Fourier transforms and classical computation.
Is it possible to use other types of unitary transformations to design new
quantum algorithms? One constraint that must be met is that the unitary transformations
must be poly-local --- they must be a product of a polynomial number of local
unitary transformations. Quantum walks on graphs might provide a good starting
point to explore
 the effects of a sequence of local unitary transformations.

The paper is organized as follows. 
We first give some background regarding classical Markov chains 
and the quantum model. We proceed to define quantum Markov chains, 
and prove various general results about the limiting distribution. 
We then prove the speed up for the quantum walk on the cycle, which 
is followed by an 
 upper bound on the mixing time for general graphs. 
Finally we prove the polynomial lower bound on the speed up for any graph,
 and conclude with a list of open questions. 

{\bf Related Work:}
Various researchers studied special cases of quantum walks on graphs.
Feynman studied quantum walks on a line;  
Farhi and Gutmann\cite{farhi} and Childs, Farhi and Gutmann\cite{childs}
studied quantum walks 
on various graphs and gave examples of graphs on which the quantum walk
hits a particular node exponentially faster than a classical walk.
(Note that this is a different task from the convergence to the stationary
distribution, which we consider in this paper.)
Ambainis, Bach, Nayak, Vishvanath and Watrous\cite{ambainis} 
studied various properties of the discrete-time 
quantum walk on the line. In particular, they have calculated the
the asymptotic behavior of the probability distributions 
for the walk on the infinite line, and shown that 
the probability distribution at time $t$ is within 
a constant in total variation distance from the uniform
distribution over an interval
which is of length linear in $t$.

\section{Background}\label{chap:background}
\subsection{Random Walks}

A simple random walk on an undirected graph $G(V,E)$, 
is described by repeated applications of 
 a stochastic matrix $P$, where $P_{u,v}=\frac{1}{d_u}$
if $(u,v)$ is an edge in $G$ and $d_u$ the degree of $u$. 
  If $G$ is connected and non-bipartite, then the distribution
 of the random walk, 
$D_t=P^t D_0$ converges to a stationary 
 distribution $\pi$ which is independent of the initial distribution $D_0$.
For $G$ which is $d-$regular, i.e. if all nodes have the same degree, 
the limiting probability distribution is uniform over the nodes of the graph. 
There are many definitions which capture  the rate of the convergence
to the limiting distribution. A survey can be found in \cite{aldous}.   

\begin{defn} {\bf Mixing Time:}  
\[ M_\epsilon=\min \{T| \, \forall t\ge T ,D_0 : \,  
||D_t-\pi|| \leq \epsilon\}, \]
\end{defn}
where here and throughout the paper, we use the total variation 
distance to measure the distance between two distributions $d_1,d_2$:  
$\|d_1-d_2\|= \sum_i |d_1(i)-d_2(i)|$.  
\begin{defn} {\bf Filling Time:}  
\[ \tau_\epsilon=
\min \{T| \, \forall t\ge T, D_0, X\subseteq V : \,  
D_t(X)\geq (1-\epsilon)\pi(X)\}. \]
\end{defn}
\begin{defn} {\bf Dispersion Time:}  
\[ \xi_\epsilon=
\min \{T| \, \forall t\ge T, D_0, X\subseteq V : \,  
D_t(X)\leq (1+\epsilon)\pi(X)\}. \]
\end{defn}

The mixing time is related to the
gap between the (unique) largest 
eigenvalue $\lambda_1 =1$ of the stochastic matrix $P$,  and the
second largest eigenvalue $\lambda_{2}$.
\begin{theo}\label{spect} {\bf Mixing time and spectral gap:} \cite{alistair}
\begin{equation}
\frac{\lambda_2}{(1-\lambda_2) \log 2 \epsilon} \leq M_\epsilon \leq \frac{1}{(1-\lambda_2)} (\max_{i} \log \pi_i^{-1}+\log \epsilon^{-1})
\end{equation}
\end{theo}

The mixing time of a random walk on a graph 
is strongly related to a geometric property
 of the graph, the {\bf conductance}, denoted by $\Phi$.

 \begin{deff} Let the {\bf capacity} $C_X$ and the {\bf flow} $F_X$ of a subset $X \subset G$ of the graph $G$ be defined as
\begin{equation}
C_X=\sum_{u \in  X} \pi_u \quad \quad F_X=\sum_{u \in X, v {\not \in} X} p_{u,v} \pi_u. 
\end{equation}
where $\pi$ is the stationary distribution, and $p_{u,v}$ is the transition 
probability. 
Then the {\bf conductance} is 
\begin{equation}
\Phi=\min_{\stackrel{0 < |X| <|G|}{C_X \leq 1/2}} \frac{F_X}{C_X}
\end{equation}
\end{deff}

\ignore{

\begin{deff} 
\begin{equation}
\Phi=\min_{\stackrel{S \subseteq V}{|E(S)| \leq  |E|/2}} \frac{|E(S,\bar{S})|}
{|E(S)|}
\end{equation} 
\end{deff}}

\begin{theo}\label{cond} {\bf Conductance and spectral gap:}[Jer\-rum, 
Sinclair\cite{jerrum}]
\begin{equation}
\frac{\Phi^2}{2} \leq (1-\lambda_2) \leq 2 \Phi 
\end{equation}
\end{theo}

Theorems \ref{spect} and \ref{cond} together imply that the mixing time 
of a Markov chain is bounded between $\Omega(1/\Phi)$ and $O(1/\Phi^2)$.

{\bf Example}
It is well known that for the
 simple random walk on an $n-$cycle, the mixing time is quadratic, 
 $M_\epsilon=\theta(n^2\cdot \log(1/\epsilon))$, and so are the filling time and the dispersion time. The conductance of this chain is $1/n$, 
which gives a lower bound of $\Omega(n)$ time steps for convergence, 
and an upper bound of $O(n^2)$. 
\ignore{
 Compare this to the bounds on $\tau_{\epsilon} $ obtained from
 the conductance: the lower bound on $\tau_{\epsilon}$ is $O(\frac{n}{\log 2 \epsilon})$, the upper bound is $O(n^2 (\log n + \log \frac{1}{\epsilon}))$.
}
\ignore{
\begin{theo} {\bf Mixing on a circle:} Let $P^{*T}$ be the probability distribution of simple random walk on the circle with $n$ (odd, $n >7$) sites after $T$ steps. Then for $T\geq n^2$, $\alpha=\pi^2 /2$ and $\beta=\pi^4 /11$
\begin{equation}
\frac{1}{2} e^{-\frac{\alpha T}{n^2} -\frac{\beta T}{n^4}} \leq ||P^{*T}-U|| \leq e^{-\frac{\alpha T}{n^2}}
\end{equation}
\end{theo}

For the circle the above implies that $\tau_\epsilon=O(n^2 \log \frac{1}{\epsilon})$. 
}

\subsection{Quantum Computation}

\noindent
{\bf The model}.
Consider a finite Hilbert space $\cal H$
with an orthonormal set of basis states
$\{\ket{s}\}$ for $s\in \Omega$. 
The states $s\in \Omega$ may be interpreted as 
the possible classical states of the system described by
$\cal H$. In general, the state of the system, $\ket{\alpha}$, 
is a unit vector in the Hilbert space $\cal H$, and can be written as 
$|\alpha\ra=\sum_{s\in \Omega} a_s |s\ra$, 
where
$\sum_{s\in \Omega} |a_s|^2=1$.
 $\la\alpha|$ denotes the conjugate transpose of $|\alpha\ra$.
 $\la\beta|\alpha\ra$ denotes the inner product of 
$|\alpha\ra$ and $|\beta\ra$.
\ignore{ (If $|\psi\ra=\sum a_s |s\ra$, 
$|\phi\ra=\sum b_s |s\ra$, then $\la\psi|\phi\ra=a^*_s b_s$.
$|\psi\ra\la\phi|$ denotes the outer product of $\psi$ and $\phi$
(an $|\Omega|\times |\Omega|$ matrix with entries $a_s b^*_s$).}
A quantum system can undergo two basic operations:
unitary evolution and measurement.

\begin{description}
\item[Unitary evolution]:
Quantum physics requires that the evolution of quantum states
is unitary, that is the state $|\alpha\ra$ is mapped to $U|\alpha\ra$, 
where  $U$ satisfies 
 $U\cdot U^\dagger=I$, and $U^\dagger$ denotes 
the transpose complex conjugate of $U$. 
 Unitary transformations
preserve norms, can be diagonalized with 
an orthonormal set of eigenvectors, 
and the corresponding eigenvalues are all of absolute value $1$. 
\item[Measurement]:
We will describe here only 
a measurement  in the orthonormal basis $|s\ra.$   
The output of the measurement of the state $|\alpha\ra$
is an element $s\in \Omega$, 
with probability $|\la s|\alpha\ra|^2$. 
Moreover, the new state of the system after the measurement is $|s\ra$. 

\item[Combining two quantum systems]:
If $\h_A$ and $\h_B$ are the Hilbert spaces of two systems, $A$ and $B$, 
then the joint system is described by the tensor product of the 
 Hilbert spaces, $\h_A\otimes \h_B$.
If the basis states for  $\h_A$, $\h_B$ are $\{|a\ra\},\{|v\ra\}$,
 respectively, 
then the basis states of $\h_A\otimes \h_B$ are $\{|a\ra\otimes |v\ra\}$. 
We use the abbreviated  notation $|a, v\ra$ for the state 
 $|a\ra\otimes |v\ra$.
This coincides with the interpretation 
by which the set of basis states of the combined system 
$A,B$ is spanned by all possible classical configurations of the 
two classical systems $A$ and $B$.   

\item[Non-unitary evolution]
The unitary model of quantum computation is not 
the most general model possible. 
In fact, the most general 
 quantum state is a semi definite positive trace one 
 matrix, $\rho$,  called the density matrix.
The density matrix of $|\alpha\ra$ is 
$|\alpha\ra\la\alpha|$. 
$\rho$ evolves by a unitary operator $U$ to 
$U\rho U^\dagger$. In general, the evolution of the density matrix 
is not necessarily unitary; 
$\rho$ evolves to $E\rho$, where $E$ is a completely positive 
linear operator, or a super operator. 
Another way to think of non unitary evolution is by adding 
qubits to the system, applying unitary transformation on the entire system 
 and then 
throwing the extra qubits away.  
For more details see \cite{AKN98,nielsen}.

\end{description}
\section{Quantum Markov Chains} \label{chap:defs}

\subsection{Definitions}\label{def}

Let $ G(V,E)$ be a graph, and let 
 ${\cal H}_V$ be the Hilbert space spanned
by states $|v\ra$ where $v\in V$. We denote by $n$, or $|V|$ 
the number of vertices in $G$. 
 First assume 
that $G$ is $d$-regular. Let ${\cal H}_A$ be an auxiliary Hilbert space
of dimension $d$ spanned by the states $\ket{1}$ through $\ket{d}$
(we think of this auxiliary Hilbert space as the ``coin space'').
Let {\bf C} be a unitary transformation on ${\cal H}_A$ (which we 
think of as the ``coin-tossing operator''). 
Label each directed edge with
a number between $1$ and $d$, such that for each $a$, the directed edges 
labeled $a$ form a permutation.
For Cayley graphs 
the labeling of a directed edge is simply 
the generator associated with the edge. 
 Now we can define a shift operator 
{\bf S} on $\cal H_A \otimes \cal H_V$ such that {\bf S}$\ket{a,v} = 
\ket{a,u}$ where $u$ is the 
$a$-th neighbor of $v$. Note that since the edge labeling is a permutation, 
 $S$ is unitary.
One step of the quantum walk is given by $U=S\cdot (C\otimes I)$.  
We call this walk a {\bf coined quantum walk.} 
{~}

\noindent{\bf Example: Coined Quantum Walk on the Cycle} 
Consider the graph 
$G$ which is a cycle with $n$ nodes. 
This $2$-regular graph can be viewed as the
 Cayley graph of the Abelian group $Z_n$ 
with the generators $+1$ (denoted by $R$ for right) and $-1$
(denoted by $L$ for left). 
The Hilbert space of the walk would then be $\C^2\otimes \C^n$.  
We choose the coin tossing operator to be the Hadamard 
transform, 
\begin{equation}
{\bf H}=\frac{1}{\sqrt{2}}\left( \begin{array}{cc} 1 & 1\\ 1 & -1 \end{array} \right)
\end{equation} 
and the shift $S$ is defined by 
\begin{eqnarray}
{\bf S}|R,i\ra&=&|R,i+1 ~mod~ n\ra\\\nonumber
{\bf S}|L,i\ra&=&|L,i-1 ~mod~ n\ra
\end{eqnarray}  
The quantum walk is then defined to be repeated applications of the Hadamard 
matrix operating on the first register, followed by the shift ${\bf S}$. 
Note that the coin we use corresponds to a classical `` unbiased''
walk,  in the sense that if measured, 
 the walk has an equal chance of moving left or right. 
{~}

In our more general definition, the {\bf general quantum walk}, 
we relax our restriction on the exact form of $U$, and 
require only that $U$ {\it respects} the structure of the graph. 
In other words,  we require that, for any $v$ and $a$,
the superposition $U|a,v\ra$ only contains
basis states $|a',v'\ra$ with $v'\in Q(v)\cup\{v\}$, 
where $Q(v)$ is the set of adjacent nodes to $v$. 
This means that the quantum walk only moves 
to neighbors of $v$ or stays at $v$.
More formally, let $X$ be a set of vertices, 
and $B$ the set of vertices which are neighbors 
of vertices in $X$ (but that are not in $X$.)
We denote by $P_X,P_B$ the probability to measure 
a vertex in $X,B$ respectively. 
\begin{claim}
\label{lbc2}
For any state $|\alpha\ra$,
\[ P_X (U|\alpha\ra)\leq P_X(|\alpha\ra) +
 P_B(|\alpha\ra) .\]
\end{claim}
\noindent
{\bf Proof:}
Let $|\alpha\ra=|\alpha_1\ra+|\alpha_2\ra$,
with $|\alpha_1\ra$ being a superposition
over vertices in $X\cup B$ and $|\alpha_2\ra$
being a superposition over $\bar X-B$.\\
Then, $P_X U|\alpha_2\ra=0$ because $|a,v\ra$ components of
$|\alpha_2\ra$ get mapped to components corresponding
to neighbors of $v$ and no vertex in $\bar X-B$ is connected by an edge
to a vertex in $X$.
Therefore, $P_X U|\alpha\ra=P_X U|\alpha_1\ra$.
Since $P_X$ is a projection (and can only decrease norm) and
$U$ is unitary, $P_X U|\alpha_1\ra \leq \|U|\alpha_1\ra\|^2
=\|\alpha_1\|^2$.
Since $|\alpha_1\ra$ is a superposition over
vertices in $X\cup B$,
$\|\alpha_1\|^2=\|P_X|\alpha_1\ra\|^2+\|P_B|\alpha_1\ra\|^2$.
$\Box$
{~}

In our most general definition, the {\bf Non Unitary Quantum Walk,}
 we allow the quantum operation 
representing one time step of the Markov chain to be non-unitary, 
i.e. the unitary matrix $U$ is replaced by 
a completely positive linear operator $E$ (a {\it super-operator})
 operating on the state of the 
system, represented by the density matrix $\rho$ 
on the Hilbert space ${\cal H}={\cal H}_G\otimes {\cal H}_A$.
We say that the walk defined by $E$ respects the graph $G$
if for any density matrix $\rho$ on ${\cal H}$
and all subsets $X$ of the vertices, 
\[P_X(E\rho)\le P_X(\rho)+P_B(\rho).\]

In the rest of the paper we use the unitary definition, 
but all definitions extend in a natural way to the non-unitary case.

\subsection{Limiting Distribution}
We now discuss the evolution of a quantum walk as a function of time. 
Starting with an initial state $|\alpha_0\ra$, 
the state of the quantum walk at time $t$ is 
$|\alpha_t\ra=U^t|\alpha_0\ra.$
In general the limit 
$
\lim_{t\mapsto\infty} |\alpha_t\ra$ 
does not exist. The reason being that $U$, as a unitary transformation, 
preserves the norm of $|\alpha_t\ra-U|\alpha_t\ra$. 
\ignore{ In fact, it exists only if 
 $|\alpha_0\ra$ is an eigenvector of $U$ with eigenvalue $1$. }
Consider instead the probability distribution on the nodes of the graph 
induced by $|\alpha_t\ra$,
\begin{deff}
$P_t(v|\alpha_0)=\sum_{a\in A}|\la a,v|\alpha_t\ra|^2$.
\end{deff}
We will sometimes denote this probability by 
$P_t^{\alpha}(v)$. 
One might ask whether this probability distribution converges to a limit.
However, $P_t$ does not converge either. 
To see this, first observe that the eigenvalues of $U$
are of the form $e^{i\theta}$, and therefore after a finite 
number of steps, $t$, $e^{i\theta t}$ is arbitrarily close to $1$
simultaneously for all eigenvalues. Hence
 the evolution of the state is quasi periodic --- the state of the system
$U^t|\alpha_0\ra$  
is arbitrarily close to $|\alpha_0\ra$ (and $U^{t+1}|\alpha_0\ra$  
is arbitrarily close to $|\alpha_1\ra$) for infinitely many times $t$.
As long as the probability distributions at time $0$ and $1$ are different, 
 $P_0 \neq P_1$, this implies that $P_t$ does not converge.

Despite the fact that the actual distribution 
does not converge, its average over time does. We define:

\begin{deff}
$\bar{P}_T(v|\alpha_0)=\frac{1}{T}\sum_{t=0}^{T-1} P_t(v|\alpha_0)$
\end{deff}
It turns out, as we will see soon, that for any initial state this 
quantity always has a limit as $T$ grows to infinity, which we denote
by $\pi(v)$ (and sometimes write $\pi(v|\alpha_0)$ if we wish to stress 
its dependence on the initial vector). Intuitively, this quantity captures 
the proportion of time which the walk ``spends'' in any given node.
Note that it is easy to sample according to this distribution 
$\bar{P}_T$ using the following process: 
Uniformly pick a random time $t$ between 
$0$ and $T-1$, let the process evolve for $t$ 
time steps and then measure to see which node it is at. 
The node will then be distributed according to $\bar{P}_T$.

We now prove a general statement about the convergence of $\bar{P}_T$.
The algebra used to prove this theorem will be useful in the rest of 
the paper. Let $|\phi_j\ra$, $\lambda_j$
 denote the  eigenvectors and corresponding 
eigenvalues of $U$, respectively. 

\begin{thm}\label{algebra}
For an initial state $|\alpha_0\ra=\sum_j a_j |\phi_j\ra$,  
\[\lim_{T\mapsto\infty} \bar{P}_T(v|\alpha_0)=
\sum_{i,j,a} a_i a_j^* \la a, v|\phi_i\ra \la \phi_j|a,v\ra \]
where the sum is only on pairs $i,j$ such that 
$\lambda_i=\lambda_j$. 
\end{thm}
\noindent
{\bf Proof:}
We start by writing down the probability 
to measure the basis state $|a,v\ra$ 
in $|\alpha_t\ra$, for a fixed $t$.  
\begin{eqnarray}
\nonumber 
|\la a,v| \alpha_t\ra|^2=|\sum_i a_i \lambda_i^t \la a,v| \phi_i\ra | ^2 \\
\label{baderech}
=\sum_{i,j} a_i a_j^* 
(\lambda_i \lambda_j^*)^t \la a,v| \phi_i\ra \la  \phi_j|a,v\ra
\end{eqnarray}
We now take the average over time of (\ref{baderech}), from $t=0$ to $T-1$. 
The only time dependent term in the above expression is 
 $(\lambda_i \lambda_j^*)^t$. 
Hence, we are interested in
\begin{equation}\label{aver}
\frac{1}{T}
\sum_{t=0}^{T-1} (\lambda_i \lambda_j^*)^t
\end{equation}
We separate into two cases: 
One in which $\lambda_i \lambda_j^*=1$, or equivalently
 $\lambda_i= \lambda_j$. 
In this case, we have that the average in Eq. 
(\ref{aver}) is equal to $1$. 
In all other cases, we can write 
\begin{equation}\label{spectralgap}
|\frac{1}{T}
\sum_{t=0}^{T-1} (\lambda_i \lambda_j^*)^t|= 
\frac{|1-(\lambda_i \lambda_j^*)^T|}{|1-\lambda_i \lambda_j^*|}|\le \frac{2}{T|\lambda_i-\lambda_j|}
\end{equation}
The latter term converges to zero, therefore the contribution 
to the limiting distribution comes solely from terms with 
$\lambda_i=\lambda_j$. 
Thus, the limiting distribution can be derived from 
the expression in equation \ref{baderech} by summing only over 
pairs which correspond to equal eigenvalues. 
This yields the desired claim, using the fact that 
the probability to measure a node $v$
 is a sum over the
probabilities to measure $|a,v\ra$, 
and so we can let each term converge separately. $\Box$ 

In the case in which all eigenvalues of $U$ are distinct, 
the limiting distribution takes a very simple form. 
Denote by $p_i(v)$ 
the probability to measure the node $v$ in the 
eigenstate $|\phi_i\ra$, so $p_i(v) =\sum_a |\la a, v |\phi_i\ra|^2$.
\begin{coro}\label{distinct}
If all eigenvalues of $U$ are distinct, then  
for an initial state $|\alpha_0\ra=\sum_j a_j |\phi_j\ra$,
\[\lim_{T\mapsto\infty} \bar{P}_T(v|\alpha_0)=\sum_{i} |a_i|^2 p_i(v).\]
\end{coro}

\ignore{
Using lemma \ref{algebra}
one can actually write the exact expression for the limiting distribution
for any quantum walk, given the initial state: 
\begin{coro} \label{limit2}
Let the initial state of the quantum walk be 
$|\alpha_0\ra= \sum_j a_j |\phi_j\ra$ (where $|\phi_j\ra$ are the eigenstates of U). 
The limiting distribution satisfies
\[ \pi(v| \alpha_0)=\sum_{i,j,a} a_i a_j^* \la a,v|\phi_i\ra \la \phi_j|a,v\ra. \]
where the sum is only on pairs $i,j$ such that 
$\lambda_i=\lambda_j$. 
If all eigenvalues of $U$ are distinct, this becomes
\[\pi_0(v|\alpha_0)=\sum_j |a_j|^2 Pr_j(v).\]
where $Pr_j(v)=\sum_a |\la v,a | \phi_j \ra|^2$ is the probability to measure the node $v$ in the $j$th eigenvector.
\end{coro} 
\noindent
{\bf Proof:} This follows immediately from lemma \ref{algebra}, 
again summing over all possible states of the auxiliary space to get 
the probability to measure a node. $\Box$}

By corollary \ref{distinct}, 
the limiting distribution depends on the initial state. 
However, if all eigenvectors induce a uniform distribution 
over the nodes of the graph, 
the limiting distribution is uniform, as is easily implied by the theorem. 
We show:

\begin{thm}\label{uniform}
Let $U$ be a coined quantum walk on the Cayley graph 
of an Abelian group, such that all eigenvalues of $U$ are distinct.
Then the limiting 
distribution $\pi$ is uniform over the nodes of the graph, 
independent of the initial state $|\alpha_0\ra$.
 \end{thm}

\noindent{\bf Proof:}
We derive an explicit expression for 
the eigenvectors of $U$, which, for a coined quantum walk, 
is of the form $U=S\cdot (C\otimes I)$. 
We note that $S$ is a matrix of dimension $dn$, for $n=|V|$.  
$S$ is composed of $d$ blocks, each of dimension $n$.
The $a-$th block  corresponds to applying the $a$-th  generator $g_a$ on the 
group. 
We note that the characters of the group, 
 $|\chi_k\ra=\frac{1}{\sqrt{n}}\sum_{v} \chi_k(v) |v\ra$, 
   are simultaneous eigenvectors
of all the blocks. The eigenvalue associated with applying 
the $a$-th block on $|\chi_k\ra$ is $\chi_k(g_a^{-1})$. 
 Since the application of the coin applies an identity on ${\cal H}_V$, 
a natural guess for the form of the eigenvectors is 
$(\sum_{a=1}^d c_a |a\ra)\otimes |\chi_k\ra$. 
Applying $C$ and then $S$ on this vector, we find that 
this vector is an eigenvector of $U$ if $\sum_{a} c_a |a\ra$ 
is an eigenvector of the $d\times d$ matrix ${\bf H}_k=\Lambda_k\cdot C$, 
where $\Lambda_k$ is a diagonal matrix, with $\Lambda_k(a,a)=
\chi_k(g_a^{-1})$. 
Since ${\bf H}_k$, as a product of two unitary matrices, has 
$d$ orthogonal eigenstates, the tensor products of these eigenstates
(which depend on $k$) 
with
$|\chi_k\ra$ give
$d$ orthonormal eigenstates for $U$. Running over $k$, 
this gives an orthonormal 
set of $nd$ eigenstates for $U$. 
It is easy to see that the probability distribution that these eigenstates
 induce on the group 
elements is uniform, since the 
 characters $|\chi_k\ra$ are uniformly distributed over the group, 
and since the eigenstates are of the form of a tensor product,
the probability to measure $a,v$ in 
$(\sum_{a=1}^d c_a |a\ra)\otimes |\chi_k\ra$ 
summed over $a$ is just the probability 
to measure $v$ in   $|\chi_k\ra$. 
This proves the theorem, using corollary  \ref{distinct}. 
 $\Box$

We claim that for any quantum walk, 
if the limiting distribution is independent of the 
initial node and state of the auxiliary space,
 then it must be uniform over the nodes. 
\begin{claim}\label{uniformlimit}
Consider a quantum walk such that for any initial basis state
of the form $|a,v\ra$,  for $v\in V$,
the limiting distribution over the nodes of the graph is equal 
to $\pi$. Then $\pi$ is uniform over the nodes of the graph. 
\end{claim}

{\bf Proof:}
If the initial state is chosen 
randomly from a uniform distribution over all basis states, 
then the limiting distribution is equal to the average over the 
limiting distributions for each initial state, 
but since they are all equal to $\pi$, the limiting distribution 
for the uniform mixture is $\pi$. However, 
the density matrix which represents a complete mixture, 
i.e. a uniformly random basis state of the space spanned by $|a,v\ra$
is preserved 
under unitary transformation, since the unitary matrix 
maps this space into itself.  Hence for any time $t$ it induces 
a uniform probability distribution over the nodes in the graph,
because the initial density matrix induces this distribution.  
This means that the limiting probability distribution starting 
from the complete mixture is uniform. Combining the two facts together, 
we get that $\pi$ is uniform. $\Box$

\subsection{Mixing Times}
We first define the analogue of the classical notion of mixing time:

\begin{deff}{\bf Mixing time:}
The mixing time  $M_\epsilon$, of a quantum Markov chain
is 
\[M_\epsilon=\min\{T|\, \forall t\ge T, |a,v\ra : \, 
\|\pi(\cdot|a,v)- \bar{P}_t(\cdot|a,v)\|\le \epsilon\}.\]   
\end{deff} 
where by the notation $P(\cdot|a,v)$ 
we mean the probability distribution conditioned 
on the initial state being $|a,v\ra$. 
This quantity measures the number of
time steps required for the average distribution to be 
$\epsilon$-close to the limiting distribution,
 starting from a basis state.

We next define a closely related quantity which we call 
sampling time: 

\begin{deff}{\bf Sampling time:}
The Sampling time  $S_\epsilon$, of a quantum Markov chain
is 
\begin{eqnarray*} 
S_\epsilon=\min\{T|\, \forall t\ge T, |a,v\ra, X\subseteq V: \\
   \, 
|\pi(X|a,v)- \bar{P}_t(X|a,v)|\le \epsilon \pi(X|a,v)\}.
\end{eqnarray*}
\end{deff} 

This is the time it takes for the walk to approximate the limiting 
distribution point-wise. 
Sampling at a random  
 time between $0$ and $S_{\epsilon}-1$ results in a distribution 
which is $\epsilon$-close point-wise to the limiting distribution,  
justifying the term {\it sampling time.}
In the same sense, sampling at a random  
 time between $0$ and $M_{\epsilon}-1$ results in a distribution 
which is $\epsilon$-close to the limiting distribution 
in total variation distance.

The third quantity,
namely the {\it filling time}
of the quantum Markov chain is defined as the first time at which 
the walk can claim to have visited all sets with at least $(1-\epsilon)$ the 
correct proportion:  
\begin{deff}{\bf Filling time:} 
The filling time,  $\tau_\epsilon$, of a quantum Markov chain
is
\begin{eqnarray*}
 \tau_\epsilon= \min \{T| \,\forall 
 X\subseteq V,|a,v\ra \,\, \exists t\le T: \\ \,\, 
P_t(X|a,v)\ge (1-\epsilon)\pi(X|a,v)\}. 
\end{eqnarray*}  
\end{deff}
\ignore{
The probabilities in the above definition are well defined, even though 
no measurement is taking place. 
At each time $t$, the state at time $t$ induces a 
probability distribution on the nodes of the graph. }

We also define the {\em dispersion time}, which is 
in some sense the opposite definition to filling time:    
\begin{deff} {\bf Dispersion time:}
The dispersion time, 
$\xi_\epsilon$, of a quantum Markov chain is
\begin{eqnarray*} 
\xi_\epsilon= \min \{T| \,\,\forall X\subseteq V,|a,v\ra \,\,
\exists t \le T :\\  \,\, P_t(X|a,v)\le (1+\epsilon)\pi(X|a,v)\}.
\end{eqnarray*}  
\end{deff}
This quantity measures how fast the quantum walk escapes any subset 
of the nodes. 
{~}

\noindent 
{\bf Remark:} 
We note that one could consider all the above definitions of mixing times 
with an arbitrary initial state, $|\alpha_0\ra$,
 and not restrict the initial state to be 
a basis state of the form $|a,v\ra$. 
However, the mixing time could change significantly. 
We will see in the cycle example that the mixing time is almost linear 
for initial basis states of the form $|a,v\ra$,
 but it is actually quadratic for general 
initial states. 

{~}

The above definitions can be related one to another 
in various ways. First, it turns out that the sampling time
is an upper bound on the mixing time, the 
filling time and the dispersion time: 

\begin{thm}\label{cezaro-is-enough}
 $ M_\epsilon, \xi_\epsilon, \tau_\epsilon\le S_\epsilon.$ 
\end{thm}
{\bf Proof:}
Fix a subset of the nodes $X$, and an initial state $|a,v\ra$. 
Suppose at all times before $S_\epsilon$, $P_t(X|a,v)< 
(1-\epsilon)\pi(X|a,v)$.
 Then the average at time $S_\epsilon$ of the probability
to measure $X$ is less than  $(1-\epsilon)\pi(X|a,v)$.
But by definition of the sampling time this is a contradiction.  
Hence there exists some time before $S_\epsilon$ at which the probability 
for the measurement outcome to be a node in $X$ is 
$P_t(X|a,v)\ge (1-\epsilon)\pi(X|a,v)$,
 and since this is true for all $X$, we have  $\tau_\epsilon\le S_\epsilon$. 
We argue in exactly the same way to prove $\xi_{\epsilon}\le S_{\epsilon}$.
The statement $M_{\epsilon} \le  S_{\epsilon}$ follows trivially 
from the definition of total variation distance. 
 $\Box$

We will later define amplified versions of these quantities, 
and find more relations between them. Let us first proceed
to give an upper bound on the mixing time $M_\epsilon$ for the quantum 
walk on the cycle.  

\section{Quantum Walk On the Cycle}
\ignore{
We consider the coined quantum walk on the 
cycle $G$ with $n$ nodes, where we restrict the discussion to odd $n$.
This $2$-regular graph can be viewed as the
 Cayley graph of the Abelian group $Z_n$ 
with the generators $+1$ and $-1$. 
We choose $C$, the coin tossing operator, to be the Hadamard 
transform, 

\begin{equation}
{\bf H}=\frac{1}{\sqrt{2}}\left( \begin{array}{cc} 1 & 1\\ 1 & -1 \end{array} \right)
\end{equation} 
Note that this coin is `` unbiased'' in the sense that, if measured, 
 it has an equal chance of moving left or right. }

In subsection \ref{def}, we have defined the coined quantum walk on the cycle. 
We restrict the discussion to cycles of an odd number of nodes $n$.
We first show that the limiting distribution for this walk is uniform. 
\begin{thm}\label{unif} 
The limiting distribution $\pi$ for 
the coined quantum walk on the $n$-cycle, with $n$ odd, and with 
the Ha\-da\-mard transform as the coin, is uniform 
on the nodes, independent of the initial state $|\alpha_0\ra$. 
\end{thm}

\noindent {\bf Proof:}
To prove that the limiting distribution is uniform,  by theorem \ref{uniform}
 it suffices to show that all eigenvalues of $U$
 are different. 
By the proof of theorem \ref{uniform}, 
the set of eigenvalues of $U$ consists of all eigenvalues
of the matrices: 
 \begin{equation}
{\bf H}_k=\left( \begin{array}{cc} \omega^k & 0 \\ 0 & \omega^{-k}
  \end{array} \right) \cdot \left( \begin{array}{cc} \frac{1}{\sqrt{2}} & \frac{1}{\sqrt{2}}\\ \frac{1}{\sqrt{2}} & -\frac{1}{\sqrt{2}} \end{array} \right)  =
\left( \begin{array}{cc} \frac{\omega^k}{\sqrt{2}} & \frac{\omega^k}{\sqrt{2}} \\ \frac{\omega^{-k}}{\sqrt{2}} & -\frac{\omega^{-k}}{\sqrt{2}}
  \end{array} \right)
\end{equation}
where $\omega=e^{\frac{2\pi i}{n}}$.
We now show that the eigenvalues of ${\bf H}_k$ are distinct. 
 The eigenvalues of ${\bf H}_k$ are the roots of the following 
quadratic equation:
\begin{equation}\label{quadratic}
\lambda^2 -i\sqrt{2}\sin(\frac{2\pi k}{n})\lambda -1=0
\end{equation}
The two solutions to this equation are of the form $e^{i\theta_k}$, 
with $\theta_k$ being one of the two solutions for the following 
equation:
\begin{equation}\label{ev}
\sin(\theta_k)=\frac{\sin(\frac{2\pi k }{n})}{\sqrt{2}}.
\end{equation}
In particular, $|\sin(\theta_k)|\le \frac{1}{\sqrt{2}}$ which means 
 that the roots of the quadratic equation \ref{quadratic}
 are confined to two regions of the unit circle,
 $\theta_k \in [-\pi/4,\pi/4]$ and 
$\theta_k \in [3\pi/4,-3\pi/4]$. 
There are two solutions for equation \ref{ev}, $\theta_{k,1}$ and 
$\theta_{k,2}$ where
$\theta_{k,1}=\pi-\theta_{k,2}$, so they lie in different
regions, and in particular, they are distinct. 
To get equality between eigenvalues
coming from different $k$'s,  
we have to have $\sin(\frac{2\pi k }{n})=\sin(\frac{2\pi k' }{n})$, 
which implies that either $k=k'$ or $k+k'=n/2$.   
The latter equation has no solutions for odd $n$, 
which implies the theorem. $\Box$

\begin{thm} \label{cyclespeed}
For the
 quantum walk on the $n$-cycle, with $n$ odd, with the Hadamard coin, we have
\[M_\epsilon \le O(\frac{n \log n} {\epsilon^3}).\]
\end{thm}
\noindent
{\bf Proof:}
We prove an upper bound on the mixing time $M_\epsilon$, 
i.e. we give an upper bound on the total variation distance between 
the average distribution $\bar{P}_T$  and the limiting distribution
$\pi$. This is done using the following lemma which holds for any 
quantum walk. 

\begin{lemm}\label{all}
Consider a general quantum walk specified by the unitary matrix $U$, 
and let $\phi_i, \lambda_i$ be the eigenvectors and corresponding 
eigenvalues of $U$, respectively. 
For any initial state $|\beta_0\ra=\sum_i a_i |\phi_i\ra$, the
total variation distance  
 between the average probability distribution 
  and the limiting 
probability distribution satisfies 
\[
\|\bar{P}_T(\cdot|\beta_0)-\pi(\cdot|\beta_0)\|\le
2\sum_{i,j, \lambda_i\not{=} \lambda_j} 
|a_i|^2
 \frac{1}{T|\lambda_i-\lambda_j|}\]
\end{lemm}
\noindent
{\bf Proof:}
We recall that in the proof
 of lemma \ref{algebra}
we have already bounded the time dependent term in the average 
probability distribution. 
From equations~(\ref{baderech}) and (\ref{spectralgap})
we have that 
\[
|\bar{P}_T(v)-\pi(v)|\leq \]
\begin{equation}
\label{oooof}
\sum_{a,i,j, \lambda_i\not{=} \lambda_j} 
|a_i|\cdot|a_j^*|\cdot| \la a,v| \phi_i\ra| \cdot|\la  \phi_j|a,v\ra|
 \frac{2}{T|\lambda_i-\lambda_j|}
\end{equation}
We now use  $|2ab|\le |a|^2+|b|^2$ twice, and summing over $v$ we get 
that $\|\bar{P}_T-\pi\|$ is at most 
\begin{equation}
\sum_{v,a,i,j, \lambda_i\not{=} \lambda_j} 
\big(\frac{|a_i|^2+|a_j^*|^2}{2}\big)\cdot\big(\frac{| \la a,v| \phi_i\ra|^2+
|\la  \phi_j|a,v\ra|^2}{2}\big)
 \frac{2}{T|\lambda_i-\lambda_j|}
\end{equation}
Summing first over all $v$ and $a$ we get the desired bound.
$\Box$
 
We observe that in that lemma, the distances  $|\lambda_i-\lambda_j|$
are of crucial importance, and they need to be large 
 for the convergence time to be small.
By the proof of theorem \ref{unif}, the 
eigenvalues are distributed in two regimes (which we will call $R$ and $R'$)
 of the complex 
unit circle,   $\theta_{k,1}\in [-\pi/4,\pi/4]=R$ and 
$\theta_{k,2}\in [3\pi/4,-3\pi/4]=R'$. 
Near the boundaries of these regimes,  
i.e. for those $\theta$'s coming from $k$'s in the vicinity of 
$n/4,3n/4$ modulo $n$, the distance between two adjacent 
eigenvalues can be of the order of $1/n^2$. 
However, we claim that the contribution of these problematic eigenvalues is 
small, and that for the rest of the eigenvalues, the distance 
is of order $1/n$. 
We fix $0<\delta<1$ (which will later be related to $\epsilon$) 
and define
\begin{eqnarray}
R_{\delta}&=&[0,(1-\delta)\frac{\pi}{2}]\cup[(1+\delta)\frac{3\pi}{2},2\pi]
\\\nonumber
R'_{\delta}&=&[(1+\delta)\frac{\pi}{2},(1-\delta)\frac{3\pi}{2}]
\end{eqnarray}
These two regimes together cover the entire  interval $[0,2\pi]$ 
except for a $2\delta$ portion of it. 
We refer to $k$ such that $\frac{2\pi k}{n}$ is in one of these regimes 
as ``$\delta$-good'', and other $k'$s are ``$\delta$-bad''. 
We also refer to eigenvectors and eigenvalues associated with 
``$\delta$-good'' 
$k$'s as ``$\delta$-good'', and similarly for ``$\delta$-bad''. 
We will later show that if the initial state for the walk 
is a basis state, 
the contribution of the bad eigenvectors is small, because 
the projection of basis states on bad eigenvectors is small. 
But first, let us restrict our attention to an initial state
which is a superposition of good eigenvectors, and consider the convergence to 
limiting distribution in this case.  
We first give a lower bound on the spacing 
between good eigenvalues.

\begin{deff}\label{delta}
\[\Delta_{\delta}=
\min_{i,j} \{|\lambda_i-\lambda_j| \,\, s.t.\,\, i\neq j\} \]
where $i,j$ run only on $\delta$-good eigenvalues. 
\end{deff}

\begin{claim}\label{evdistances}
For the quantum walk on the odd $n$ cycle with the Hadamard coin,
$\Delta_{\delta}\ge \frac{\pi \delta}{\sqrt{2}n}$.  
\end{claim}
\noindent
{\bf Proof:}
First observe that if $\lambda_i,\lambda_j$  originate from the same $k$,
then they lie in two different regimes $R$ and $R'$, which means that  
 $|\lambda_i-\lambda_j|$ is at least $\sqrt{2}$. Hence, we can restrict 
our attention to eigenvalues coming from different $k$'s. 
Let  $\lambda_i$, $\lambda_j$ originate from $k,k'$, respectively. 
Then  using equation \ref{ev} we have
\begin{equation}\label{lower}
|\lambda_i-\lambda_j|\ge |\sin(\theta_i)-\sin(\theta_j)|
=\frac{1}{\sqrt{2}}
|\sin(\frac{2\pi k}{n})-\sin(\frac{2\pi k'}{n})|
\end{equation}
We separate the proof to two cases. In the first case, 
$k,k'$ lie in the same regime, $R_{\delta}$ or $R'_{\delta}$.  
Recall the intermediate value theorem, which states 
that for a continuous function,  for any $x\le y$, there exists 
$x\le z\le y$ such that
$
|f(x)-f(y)|=
|f'(z) (x-y)|.$
Applying this theorem with $f(x)=\sin(x)$, we get
\begin{equation}
|\sin(\frac{2\pi k}{n})-\sin(\frac{2\pi k'}{n})|
= |\cos(\gamma)(\frac{2\pi (k-k')}{n})|
\ge |\cos(\gamma)\frac{2\pi}{n}|
\end{equation}
for some $\frac{2\pi k}{n}\le \gamma \le \frac{2\pi k'}{n}$. 
Since $k,k'$ are in the same regime, 
then $\gamma\in R_{\delta}$ or $\gamma\in R'_{\delta}$,
 and by monotonicity of the
$\cos$ function, we have:   
\begin{equation}\label{value}
|\cos(\gamma)|\ge 
|\cos(\frac{\pi(1-\delta)}{2})|=|\sin(\frac{\delta\pi}{2})|\ge \delta 
\end{equation}
 where the last equality follows from the fact that 
$\sin(0)=0, \sin(\pi/2)=1$, and $\sin$ is convex in the regime 
$[0,\pi/2]$. 
If $k,k'$ belong to different regimes, 
then we can no longer claim that $\cos(\gamma)$ is large. Instead, we write 
\begin{equation}
|\sin(\frac{2\pi k}{n})-\sin(\frac{2\pi k'}{n})|=
|\sin(\frac{2\pi k}{n})-\sin(\pi- \frac{2\pi k'}{n})|
\end{equation}
\[
\ge |\cos(\gamma')(\frac{2\pi (k+k')}{n}-\pi)|\ge |\cos(\gamma')\frac{\pi}{n}|\]
for some $\gamma'$, between $\frac{2\pi k}{n}$ and $\pi-\frac{2\pi k'}{n}$. 
 Now, $\frac{2\pi k}{n}$ and $\pi-\frac{2\pi k'}{n}$ lie in the same regime, 
and so using the same argument as before, the lemma follows. $\Box$

We can now use claim \ref{evdistances}
to give a better lower bound on the distance between two eigenvalues. 
\begin{claim}\label{dist}
Let us order the eigenvalues such that 
$0 \leq Arg(\lambda_1)\le Arg(\lambda_2)....\le Arg(\lambda_{2n})\le 2\pi$.
Consider $\lambda_i$ and $\lambda_j$ which lie in the same regime, 
$R_{\delta}$ or $R'_{\delta}$. Then 
\[|\lambda_i-\lambda_j|\ge \frac{2\sqrt{2}}{\pi}|i-j|\Delta_{\delta}\]
\end{claim}
\noindent
{\bf Proof:}
Consider $i,j$ as in the requirements of the claim. 
Let $L_{i,j}$
 be the length of the shorter arc on the unit circle that connects 
$\lambda_j$ to $\lambda_i$. We first claim that 
$|\lambda_i-\lambda_j|\ge \frac{2\sqrt{2}}{\pi}L_{i,j}$ in our regime. 
This is true since the ratio $|\lambda_i-\lambda_j|/L_{i,j}$
 is monotonically decreasing
 in $\theta_i-\theta_j$ in the regime $\theta_i-\theta_j\in [0,\pi/2]$, and so we can bound 
$|\lambda_i-\lambda_j|/L_{i,j}$ from below by its value on 
 the boundary,  
 $|\theta_i-\theta_j|=\pi/2$,  which gives
$|\lambda_i-\lambda_j|/L_{i,j}=\frac{2\sqrt{2}}{\pi}$.
Hence, to bound $|\lambda_i-\lambda_j|$ we give a lower bound on $L_{i,j}$.
We have $L_{i,j}=|i-j|L_{i,i+1}$, so it suffices to bound $L_{i,i+1}$. 
This is done by noticing that
 $L_{i,j}\ge 2 |\sin L_{i,j}/2|=|\lambda_i - \lambda_j|\ge \Delta_{\delta}$,
where the first inequality uses $\sin(x)\le x$, the second equality 
uses simple trigonometry, and the last inequality uses claim 
\ref{evdistances}. 
Hence, $L_{i,j}=|i-j|L_{i,i+1}\ge |i-j|\Delta_{\delta}$ which combined with 
$|\lambda_i-\lambda_j|\ge \frac{2\sqrt{2}}{\pi}L_{i,j}$ gives the claim. 
$\Box$

\begin{claim}\label{bou}
Let $|\beta\ra=\sum_{i} a_i |\phi_i\ra$ such that all coefficients 
of $\delta$ bad eigenvectors are zero. 
Then 
\[
\|\bar{P}^{\beta}_T-\pi^{\beta}\|\le  \frac{2n (ln(n)+2)}{ T \delta}\]
\end{claim}
\noindent
{\bf Proof:}
Using lemma \ref{all} we write
\begin{equation}\label{eq:bou}
\|\bar{P}^{\beta}_T-\pi^{\beta}\|\le
 2\sum_{k}\sum'_{i,j, |i-j|=k} |a_i|^2
 \frac{\pi}{Tk 2\sqrt{2} \Delta_{\delta}} + 2\sum^{''}_{i,j} 
|a_i|^2
 \frac{1}{T\sqrt{2}}
\end{equation}
where in the first sum the prime indicates the fact that we sum
 over pairs $i,j$ in the same regime $R_\delta$ or $R'_\delta$, such that 
$\lambda_i\ne \lambda_j$, and in the second sum the double prime indicates 
that the sum is over 
 pairs $i,j$  such that $\lambda_i,\lambda_j$ are in the different regimes.
We have used  claim \ref{dist} in the first sum,  
and the fact that $|\lambda_i-\lambda_j|\ge \sqrt{2}$ 
for eigenvalues from different regimes in the second sum.

To bound  the first term, we observe that for each $i$ there are at most two eigenvalues
$j$ such that $|i-j|=k$. 
We first sum over  $i,j$. Then, summing over $k$, 
we use the fact that the sum of the first $n$ terms 
of the harmonic series is less than
 $\ln(n)+1$. Thus, the first term is at most
 $\frac{\sqrt{2}\pi (\ln (n)+1)}{T\Delta_{\delta}}$. 
For the second term we get an upper bound 
$\frac{\sqrt{2}n}{T}$. 
Using claim \ref{evdistances} to bound $\Delta_{\delta}$ in the first term we get the 
desired claim.
$\Box$

We now prove that the contribution of the $\delta$-bad vectors is small. 
This follows from the following two claims.  
\begin{claim}\label{proj}
The projection of any basis state on the bad eigenvectors 
is of norm squared at most $2\delta$. 
\end{claim}

\noindent{\bf Proof}:
The $2n$ dimensional Hilbert space of the quantum walk 
can be viewed as a direct sum of the two dimensional subspaces $L_k$, 
where $L_k$ is the space spanned by the two eigenvectors originating from 
$k$. The projection of a basis state on $L_k$ is of norm squared 
exactly $\frac{1}{n}$.  The claim follows from the fact that 
 there are $2\delta n$ bad $k's$. $\Box$

\begin{claim}\label{triangle}
Consider two initial states, $|\alpha_0\ra, |\beta_0\ra$. 
Denote by $\bar{P_T}^{\alpha}$,$\bar{P_T}^{\beta}$ the average 
distributions in the quantum walk starting with  $|\alpha_0\ra, |\beta_0\ra$, 
respectively. Then for all $T$, the total variation distance between 
the average distribution is bounded by the distance between the initial 
states: 
\[\|\bar{P_T}^{\alpha}-\bar{P_T}^{\beta}\|\le 
2\||\alpha_0\ra- |\beta_0\ra\|\]
\end{claim}

\noindent{\bf Proof}:
Denote by $|\alpha_t\ra, |\beta_t\ra$ the states at time $t$ starting with 
 $|\alpha_0\ra, |\beta_0\ra$ as initial states. 
Denote by $P_t^{\alpha},P_t^{\beta}$ the induced distributions of 
$|\alpha_t\ra, |\beta_t\ra$ on the nodes of the graph. 
Clearly, $ \|\bar{P_T}^{\alpha}-\bar{P_T}^{\beta}\|\le \max_{t\le T} 
\|P_t^{\alpha}-P_t^{\beta}\|$. 
Due to unitarity of the walk, the distance is preserved: 
$\||\alpha_t\ra-|\beta_t\ra\|=\||\alpha_0\ra-|\beta_0\ra\|$. 
By lemma $11$ in \cite{AKN98}, the total variation distance between 
the two probability distributions resulting
from a measurement on two states which are $\epsilon$ apart, 
is at most $2\epsilon$. This proves the claim. $\Box$

\begin{claim}\label{neg}
Let $|\alpha_0\ra$ be the initial basis 
state, and $|\beta_0\ra$ be the initial basis state projected 
on the $\delta-$good eigenvectors, and renormalized. Then 
\[
\|\bar{P}^{\alpha}_T-\pi^{\alpha}\|
\le 8\sqrt{2\delta}+
\|\bar{P}^{\beta}_T-\pi^{\beta}\|\]
\end{claim}
\noindent
{\bf Proof:}
We write 
\begin{equation}
\|\bar{P}^{\alpha}_T-\pi^{\alpha}\|\le 
\|\bar{P}^{\alpha}_T-\bar{P}_T^{\beta}\|+\|\bar{P}^{\beta}_T-\pi^{\beta}\|+
\|\pi^{\beta}-\pi^{\alpha}\|.
\end{equation}
The first term, by claim \ref{triangle} is smaller than 
$2\||\alpha_0\ra-|\beta_0\ra\|$. The last term is also smaller than 
$2\||\alpha_0\ra-|\beta_0\ra\|$, since it is the limit of distances which are smaller than this term. 
We claim that 
$\||\alpha_0\ra-|\beta_0\ra\|\le 2\sqrt{2\delta}$. 
This is true since by claim \ref{proj} 
we can write $|\alpha_0\ra=a|\beta_0\ra+|v\ra$,
where $|v\ra$ is a vector of norm at most $\sqrt{2\delta}$ and $a$ 
is larger than $\sqrt{1-2\delta}$.  
  $\||\alpha_0\ra-|\beta_0\ra\|\le |1-a|+\sqrt{2\delta}\le 2\sqrt{2\delta}$. 
$\Box$

{~}

We can now combine claim \ref{neg} and claim \ref{bou}
 to finish the proof of the Theorem. 
If we now pick $\delta=\frac{1}{2}(\epsilon/16)^2$, and 
$T\ge 4n (ln(n)+2)/\epsilon\delta$, 
we get that 
\begin{equation}
\|\bar{P}^{\alpha}_T-\pi\|\le 8\sqrt{2\delta}+ 
\frac{2n (ln(n)+2)}{ T \delta}\le \frac{\epsilon}{2}+\frac{\epsilon}{2}=\epsilon. ~~~~\Box   
\end{equation}

{~}

\noindent
{\bf Remark}
In the theory of classical Markov chains, the distance between 
the first and second eigenvalues plays a crucial role in mixing time 
analysis. In the quantum case, we see that the distances between eigenvalues 
play a similarly important role; 
However, unlike in the classical case, since all
 eigenvalues of a unitary matrix 
are of absolute value $1$, there is no special eigenvector which plays 
the role of the fixed state, and all eigenvalues play are equally important.

\section{Amplification}
In classical Markov chains, after approaching a certain closeness 
to the limiting distribution, the distance to the limiting distribution 
starts to drop exponentially. 
Theorem \ref{cyclespeed} gives only polynomial dependence on $1/\epsilon$
in the quantum case.  
However, one can amplify the closeness 
in a very simple way. 
Suppose the limiting distribution $\pi(\cdot|a,v)$
 is independent of the initial node $v$ and the state $a$, and is equal 
to $\pi$. 
(Recall that by claim \ref{uniformlimit} $\pi$ is uniform.)   
In this case, the closeness to $\pi$ 
can be amplified in a standard way to get logarithmic dependence on 
$1/\epsilon$. This is done by running the walk for 
 $M_{\epsilon}$ steps (i.e.  for a random time between 
$0$ and $M_\epsilon-1$) and then measuring the node.
If the measured node is $v$, we 
then initialize the state to be $|a,v\ra$
with a random auxiliary state $a$, and start the walk again for 
one more stage of $M_\epsilon$ steps, and so on for $k$ times. 
 We claim:
\begin{lemma}\label{amplification} {\bf Amplification lemma}
Running the quantum walk for $k$ amplification 
steps, each lasting $M_{\epsilon}$ time steps, 
results in a distribution which is $\epsilon^k$ close to $\pi$.
\end{lemma}

{\bf Proof:} 
Define $P_{v,u}$ to be the probability to measure the node $u$ 
starting from a random initial basis state $|a,v\ra$, 
 in one amplification step, where $a$
 is randomly chosen from all basis states of the auxiliary space.  
The matrix $P$ defined by these transition probabilities is a stochastic 
matrix.  
We claim that applying one amplification step starting from
the uniform distribution $\pi$ 
 one gets the distribution $\pi$ again. The reason is that  
 a uniformly random state $|a,v\ra$ (which induces a uniform distribution 
over the nodes $v$) is a complete mixture 
of the Hilbert space in which the walk evolves. 
The unitary transformation associated with the walk is a map from 
this space to itself, 
therefore, starting from a complete mixture of this space,  
the state of the system remains a complete mixture, i.e. $\pi$ is 
preserved.

We now claim that the $L_1$ norm of any vector 
orthogonal to $\pi$ is shrunk by a factor of $\epsilon$ by the matrix $P$. 
To prove that $\|\tau P\|\le \epsilon \|\tau\|$ for $\tau\perp \pi$, observe 
that by definition of $M_\epsilon$, for any distribution $\sigma=\pi+\tau$, 
$\|\sigma P-\pi\|\le \epsilon$. This means that for any vector $\tau$ 
for which the sum of elements is zero, and each coordinate is 
at least $-1/n$, we have $\|\tau P\|\le \epsilon$. 
We can define a basis for the subspace orthogonal to $\pi$,
which is composed of such vectors: These will be the vectors
$v_i$, where $v_i$ has $(n-1)/n$ on its $i^{th}$
 coordinate and the rest are all equal to 
$-1/n$. Any $\tau\perp \pi$ can be written as a sum of $v_i$ and $\pi$:
$\tau=\sum_i \tau_i (v_i+\pi)=\sum_i \tau_i v_i$. 
$\|\tau P\|\le \sum_i |\tau_i| \|v_i P\|\le \sum_i |\tau_i| \epsilon=
\|\tau\|\epsilon$.

We can now prove the claim by induction.
Starting from a distribution $\sigma$ which is within $\delta$ total 
variation distance from $\pi$, we can
 write $\sigma =\pi+\tau$
 where $\tau\perp \pi$, and
$\|\tau\|\le \delta$.  
Then,   $\|\sigma P-\pi\|=\|\pi P+ \tau P-\pi \|= \|\tau P\|\le \|\tau\|\epsilon\le \delta\epsilon.$ 
$\Box$

For the cases in which the limiting distribution is independent of 
the initial state, 
we can now generalize our measures of convergence to allow the possibility 
of amplification. This means that in all definitions we allow a warm
start, i.e. we first amplify for several amplification steps,
which all together last $T_A$ time steps,  to get 
an initial ``warm'' node (the exact times at which 
one measures are chosen so that $T_A$ is minimized, and the times at which 
the node is measured during $T_A$ are referred to as the 
``amplification scheme'').
Then we apply the various definitions 
of mixing times with the ``warm start''  
node as the initial node. However,   
to account for the initial amplification stage, 
 we add $T_A$ to the mixing times. 
We denote those amplified versions of convergence 
with primes. 
$M'_\epsilon$, $S'_\epsilon$, and so on. We have: 

\begin{thm}\label{primes}
$\tau'_\epsilon,\xi'_\epsilon\le S'_\epsilon \le 
\frac{\log (1/\min_v \{\pi(v)\})}{log(1/\epsilon)} M_\epsilon$
\end{thm}
{\bf Proof:}
We first prove the right inequality. 
Closeness to within $\epsilon$ point wise will be guaranteed 
if  the total variation distance is at most $\epsilon \min_v \{\pi(v)\}$. 
For that, by the amplification lemma, it suffices to apply 
$\log (1/\min \pi(v))/log(1/\epsilon)$ amplification steps, each of length  
$M_\epsilon$. The proof of the left inequality
is as follows. Let $S'_\epsilon$ be achieved with a certain 
amplification scheme. We then use the same amplification scheme 
for the dispersion time and the filling time, so that we start 
with the same distribution over initial nodes. 
Now, the remaining of the proof is exactly 
as the proof of theorem \ref{cezaro-is-enough}, 
referring only to the time interval starting at the end 
of the amplification stage. 
$\Box$ 

\begin{thm}
For the
 quantum walk on the $n$-cycle, with $n$ odd, with the Hadamard coin, we have
 \[M'_\epsilon\le O\big(n \log n log \big(\frac{1}{\epsilon}\big)\big)~,~
 S'_\epsilon,\xi'_\epsilon,\tau'_\epsilon \le O\big(n \log^2 n \log\big(\frac{1}{
\epsilon}\big)\big) \]
\end{thm}
\noindent
{\bf proof:}
The upper bound on $S'_\epsilon, \tau'_{\epsilon}$ and $\xi'_{\epsilon}$
follows from theorems \ref{primes} and \ref{cyclespeed},
 and the fact that 
$\min\{\pi(v)\}=1/n$. 
$\Box$

\section{General Graphs}
We now prove a general upper bound on $M_\epsilon$ 
for any quantum walk. This will imply upper bounds 
on the other mixing times by theorem \ref{primes}.

Let $|\phi_i\ra$ be the eigenvectors of $U$ with eigenvalues $\lambda_i$.
The upper bound will be given in terms of $\Delta$,
which is defined to be the minimal spacing between the
 eigenvalues.  
\begin{thm}
Consider a general quantum walk on a graph $G$ with $n$ nodes, 
 with an auxiliary space of dimension $d$. 
Then, for an initial state $|\alpha\ra$, 
 the total variation distance from 
its limiting distribution is  
\[\|\bar{P}_T(\cdot|\alpha)-\pi(\cdot|\alpha)\|\le 
\frac{\pi (\ln (nd/2)+1)}{ T \Delta}\]
\end{thm}
\noindent
{\bf Proof:} The proof follows approximately
the same lines as the 
proof for the upper bound for the cycle, except that the complications 
due to throwing away part of the system disappear. 
More precisely, the proof goes along the lines of the proof of claim 
\ref{bou}.
The main difference is that we do not have 
a partition of the eigenvalues into two regimes. 
For this reason, in the counterpart of claim \ref{dist},
we can have $|\theta_i-\theta_j|\in[0, \pi]$ (instead of
$[0, \pi/2]$). Then, $|\lambda_i-\lambda_j|/L_{i, j}$ is
minimized by $|\theta_i-\theta_j|=\pi$ (instead of $\pi/2$)
and we get $|\lambda_i-\lambda_j|\geq \frac{2}{\pi}|i-j| \Delta$.
Also, the counterpart of equation \ref{eq:bou} has just one summation
(over all eigenvalues) instead of two (over eigenvalues
in the same regime and eigenvalues in different regimes).
After that, we just notice that in the general case 
there are $nd$ eigenvalues, which implies  that $k=|i-j|$
runs up to $nd/2.$ 
$\Box$.

Just like we did in the cycle case, one can separate the eigenvectors to 
``good'' and ``bad'' vectors, where the proportion of the ``bad'' vectors 
is $\delta$, to get a better estimation of the mixing time.

\section{The Lower bound}

Here we are going to prove a lower bound on the various 
mixing measures of a general quantum random walk. In analogy to the classical case this bound will be stated in terms of the {\em conductance} $\Phi$ of the underlying graph $G$ (cf. Chapter \ref{chap:background}).\\ We will define a slightly different quantity $\Phi'$ first:
Let $(X, \bar X)$ be a cut in the graph $G$ (i.e., a partition of
vertices into two sets). 
Define $B_X$, the boundary of $X$, as the set of vertices
in $\bar X$ that have an edge going to $X$.
Let $\Phi'=\min_{0<|X|\leq\frac{1}{2}|V|} \frac{|B_X|}{|X|}$.

\begin{thm}
\label{lbt1}
The filling, dispersion, mixing and sampling times of a general
quantum walk with a uniform limiting distribution are $\Omega(1/\Phi')$.
\end{thm}
\noindent
{\bf Proof:}
Let $(X, \bar X)$ be the cut that achieves the minimum 
$({|B_X|}/{|X|})$. To simplify the notation, let $B=B_X$.\\
Let $\h_X$ be the Hilbert space supported by nodes in $X$, i.e. 
the subspace spanned by $\{|a,v\ra\}$ for all $v\in X$ and all basis 
states $a$ of the 
auxiliary space.  
Let $\h_{\bar X}$ and $\h_B$ be Hilbert spaces supported by nodes in $\bar X$
and $B$ (defined similarly).
We show a lower bound on 
filling and dispersion times
by taking a random state $|\alpha\ra$ of form $|a,v\ra$, $v\in \bar X$ 
and showing that the projection of $U^t|\alpha\ra$ onto $\h_X$ is small 
for all $t\le \Theta(1/\Phi')$.\\
For any state 
$|\phi\ra$, let $P_X|\phi\ra$ and $P_B|\phi\ra$ be the projections
of $|\phi \ra$ onto $\h_X$ and $\h_B$, respectively.\\
Let $|\alpha\ra$ be a uniformly random basis state $|a,v\ra$,
$v\in \bar X$. 
We bound the expected projection of $|\alpha\ra$ onto the boundary $\h_B$
and then use that to bound the expected projection
of $U^t|\alpha\ra$ onto $\h_X$. (This works because the only way 
to go from $\bar X$ to $X$ is through the boundary $B$.)

\begin{claim}
\label{lbc1}
For any $t$, the expected value of $\|P_B U^t|\alpha\ra\|^2$ is 
at most ${|B|}/{|\bar X|}$.
\end{claim}
\noindent
{\bf Proof:}
$|\alpha\ra$ is a uniformly random state in $|\bar X|d$ dimensions.
Since $U$ is unitary, $U^t|\alpha\ra$ is a uniformly random state in some
$|\bar X|d$-dimensional subspace of $\h$.
The projection of this state to the $|B|d$-dimensional 
subspace $\h_B$ is at most $\frac{|B| d}{|\bar X| d}=\frac{|B|}{|\bar X|}$
(with equality if and only if $\h_B\subseteq U^t(\h_{\bar X})$).
$\Box$


By applying Claim \ref{lbc2} to $P_X U^{k}|\alpha\ra$, we get
\[ \|P_X U^k|\alpha\ra\|^2\leq \|P_X U^{k-1}|\alpha\ra\|^2 + 
 \|P_B U^{k-1}|\alpha\ra\|^2 .\]
By applying Claim \ref{lbc2} $k-1$ more times 
(to $\|P_X U^{k-1}|\alpha\ra\|^2$,
then $\|P_X U^{k-2}|\alpha\ra\|^2$ and so on), we get
\[ \|P_X U^k |\alpha\ra\|^2 \leq \|P_B U^{k-1} |\alpha\ra\|^2 +
  \ldots + \|P_B U|\alpha\ra\|^2+ 
 \|P_B|\alpha\ra\|^2 
.\]
By claim \ref{lbc1}, the expected value of each term on the 
right-hand side (for a random $|\alpha\ra=|a,v\ra$, $v\in \bar X$) 
is at most ${|B|}/{|\bar X|}$.
Therefore, the expected value of the sum on the right hand side is
at most $k ({|B|}/{|\bar X|})$.
For some $|\alpha\ra=|a,v\ra$, 
the sum $\sum_{i=0}^{k-1} \|P_B U^i|\alpha\ra\|^2$
is at most its expectation. 
For this $|\alpha\ra$, we have
\[ \|P_X U^j|\alpha\ra\|^2 \leq \sum_{i=0}^{j-1} \|P_B U^i|\alpha\ra\|^2 
\leq \sum_{i=0}^{k-1} \|P_B U^i|\alpha\ra\|^2 \leq k\frac{|B|}{|\bar X|} \]
for all $j<k$.
For $k$ to be the filling time, one of $\|P_X U^j|\alpha\ra\|^2$ should be
at least $(1-\epsilon)\frac{|X|}{|V|}$.
Then, $k\frac{|B|}{|\bar X|}\geq (1-\epsilon)\frac{|X|}{|V|}$
and $k\geq (1-\epsilon) \frac{|X| |\bar X|}{|B| |V|}$.
Since $|X|\leq \frac{1}{2}|V|$,
$|\bar X|\geq \frac{1}{2}|V|$ and $k = \Omega(\frac{|X|}{|B|})=
\Omega(1/\Phi')$. 
A similar argument applies to dispersion time and sampling time.
The bound on mixing time is implied by theorem \ref{cezaro-is-enough}
$\Box$

The quantity $\Phi'$ (which we call {\em boundary}) is similar 
but not identical to the conductance.

\begin{lemma}\label{lbt2}
For a graph with maximal degree $d$, 
\[ \Phi' \leq d\Phi, \]
where $\Phi$ is the conductance of a simple random walk on the graph.   
\end{lemma}

\noindent{\bf Proof:}
For a simple random walk on $G$, the limiting distribution is 
$\pi_v= d_v/\sum_v d_v=d_v/2|E|$. 
Fix a cut $X,\bar{X}$, and denote the conductance of this cut 
by $\Phi_X$. 
The capacity of $X$ is 
$C_X=\frac{\sum_{v\in X} d_v}{2|E|}\le\frac{d|X|}{2|E|}$. 
Let $E(X:\bar{X})$ be the set of edges going between $\bar X$ and $X$.
The flow $F_X$ satisfies  $F_X=\frac{|E(X:\bar{X})|}{2|E|}\ge \frac{|B_X|}{2|E|}$. 
Therefore, $\Phi_X=\frac{F_X}{C_X}\ge \frac{2|B_X||E|}{2d|E||X|}\ge \frac{|B_X|}{d|X|}=\frac{\Phi'_X}{d}$. This is true for any cut $X$, 
which implies the lemma. 
$\Box$

Therefore, $\Omega(\frac{1}{\Phi'})=\Omega(\frac{1}{d\Phi})$
and theorem \ref{lbt1} implies an $\Omega(\frac{1}{d\Phi})$
lower bound on
For constant degree $d$ graphs, this lower bound is 
$\Omega(\frac{1}{\Phi})$, the same as the classical lower bound
on filling, dispersion and sampling times.
Since a classical random walk converges in 
$O(\frac{1}{\Phi^2})$ steps, this means that a quantum walk
can be at most quadratically faster.

\begin{coro}
For a general quantum walk on a bounded degree graph,  the
 filling, dispersion, sampling and mixing times are at most quadratically
faster than the mixing time of the simple classical 
 random walk on that graph. 
\end{coro}

For unbounded $d$, the factor-$d$ gap between 
the two lower bounds (quantum and classical) is important. 
This gap can be quite large: we did not rule out the case 
in which  
the quantum filling time is $O(\log^c n)$ but $d$ is $\Theta(n)$.
We suspect that the bound can be improved and 
quantum walks are at most quadratically faster on any graph.
We can prove that for the special case of coined quantum walks.

\begin{thm}
\label{lbt3}
For a coined quantum walk,
the filling, dispersion and sampling times
are $\Omega(1/\Phi)$.
\end{thm}
\noindent
{\bf Proof:}
To simplify the proof, we assume that 
the unitary transformation
$U$ is of form $C\circ S$, not $S\circ C$
(i.e. we first do the shift $S$ and then the coin flip $C$). 
This assumption can be removed by replacing the starting state 
$|\alpha\ra$ by $C^{-1}|\alpha\ra$ and
adding an extra $C$ at the end.\\
The proof is similar to Theorem \ref{lbt1}.
We take the set $X$ which achieves the conductance, 
 $\Phi=\frac{F_X}{C_X}$.
Let $\h_X$ and $\h_{\bar{X}}$ be similar to the proof of Theorem \ref{lbt1}
and $\h_C$ be the Hilbert space spanned by {\em edges} in the cut
(i.e., the space spanned by $|b,v\ra$, $v\in \bar X$, $b \circ v\in X$).
Let $P_X$ and $P_C$ be the projections onto $\h_X$ and $\h_C$, respectively.

\begin{claim}
\label{lbc3}
For any $t$, the expected value of $\|P_C U^t|\alpha\ra\|^2$ (for a 
uniformly random $|\alpha\ra=|a,v\ra$, $v\in \bar X$) is 
at most $\frac{|E(X:\bar{X})|}{|\bar X|d}$.
\end{claim}
\noindent
{\bf Proof:}
$|\alpha\ra$ is a uniformly random state in $|\bar X|d$ dimensions.
Since $U$ is unitary, $U^t|\alpha\ra$ is a uniformly random state in some
$|\bar X|d$-dimensional subspace of $\h$.
The projection of this state to the $|E(X:\bar{X})|$-dimensional 
subspace $\h_C$ is at most $\frac{|E(X:\bar{X})|}{|\bar X| d}$.
$\Box$

\begin{claim}
\label{lbc4}
For any state $|\alpha\ra$ and any $k\in\Nn$, 
\[ \|P_X U^k|\alpha\ra\|^2\leq \|P_X U^{k-1}|\alpha\ra\|^2 + 
 \|P_C U^{k-1}|\alpha\ra\|^2 .\]
\end{claim}
\noindent
{\bf Proof:}
Define $|\alpha'\ra=U^{k-1}|\alpha\ra$.
Let $|\alpha'\ra=|\alpha'_1\ra+|\alpha'_2\ra$, with $|\alpha'_1\ra$ being a 
superposition over $|b,v\ra$ with $v\in X$ or $b\circ v\in X$
and $|\alpha'_2\ra$ being a superposition over all other $|b,v\ra$.\\
Then, $S|\alpha'_2\ra\in \h_{\bar X}$ because $b\circ v\notin X$ 
for all $|b,v\ra$ that appear in $|\alpha'_2\ra$.
Since the coin flip $C$ does not change $v$, 
this also means that $U|\alpha'_2\ra=C\cdot S|\alpha'_2\ra \in \h_{\bar X}$.
Therefore,
$P_X U|\alpha'\ra= P_X U|\alpha'_1\ra$.\\
Similarly to the proof of claim \ref{lbc2},
$ \|P_X U|\alpha'_1\ra\|^2 \leq \||\alpha'_1\ra\|^2 =
\|P_X|\alpha'_1\ra\|^2 + \|P_C|\alpha'_1\ra\|^2 =
\|P_X|\alpha'\ra\|^2 + \|P_C|\alpha'\ra\|^2.$
$\Box$

The rest of proof is identical to Theorem \ref{lbt1}.
$\Box$

\subsection{Lower bound for non-unitary walks}
We first deal with a  special case of non-unitary walks. 
In this case, instead of $U$, we have a set of possible unitary 
matrices $U_i$, and we choose one of them randomly to apply 
at time $t$. The lower bounds extend trivially to this case.

We now give a simple lower bound on the sampling time of 
a general non-unitary walk in terms of the boundary. 
\begin{theo}
The sampling time of a general non-unitary quantum walk 
with a uniform limiting distribution are $\Omega(\frac{1}{\Phi'})$.
\end{theo}
{\bf Proof:} 
Fix a cut, $X,\bar{X}$.
Suppose we start with a state concentrated on $\bar{X}$.  
By applying the definition of a non-unitary quantum walk that
respects the structure of the graph (section \ref{def})
several times we get
\begin{equation}
 P_X E^k \rho \leq
 P_B E^{k-1} \rho + \ldots + P_B E \rho+ 
P_B\rho 
\end{equation}
For $S_\epsilon$ to be the sampling
 time, there must be some $S_\epsilon\le k\le 2S_\epsilon$ such that 
$P_X E^k\rho\ge (1-3\epsilon)\frac{|X|}{|G|}$. 
On the other hand, the sum at the 
right hand side of the above equation is exactly 
$k\bar{P}_k(B|\rho)$, which, for $k> S_\epsilon$, 
must satisfy
\begin{equation}
\bar{P}_k(B|\rho)\le (1+\epsilon)|B|/|G|
\end{equation}
This means that 
$(1-3\epsilon)\frac{|X|}{|G|}\le k(1+\epsilon)|B|/|G|$
or 
\begin{equation}
\frac{(1-3\epsilon)|X|}{2(1+\epsilon)|B|}\le S_\epsilon
\end{equation}
for all $X$. 
If we pick $X$ to be the set which achieves the minimum boundary
$|B|/|X|$, we get the desired result. 
$\Box$

We leave it as an open question to generalize the lower bound 
for non-unitary walks to other mixing measures. 

\section{Concluding Remarks}
In this paper we have set up the basic definitions for 
quantum walks on graphs. However,  
the foundations of the theory of quantum walks on graphs still 
await discovery. We list here a few selected
open problems. 

The first open question is for which graphs quantum speed up
is achievable. More generally, can the $1/\Phi$ lower bound 
 always be achieved quantumly? 
In \cite{lifting} it was shown that for any Markov chain, 
there exists a lifted version of it which achieves this bound, 
but no lifting can give better than  $1/\Phi$ convergence. 
To achieve the lower bound of  $1/\Phi$ by lifting, 
one has to be able to solve the multi-commodity flow on the graph, 
a task which is in general extremely hard. Therefore 
the lifting speed-up is an existence proof, rather than 
an algorithmic one. It would therefore be very interesting 
to know whether  convergence in time $1/\Phi$ 
 can be achieved by quantum walks for graphs 
other than the cycle  
in an efficiently constructible way.

An open question is to make our two bounds tight.   
We have shown how to improve the factor of $1/d$ in the lower bound 
for coined quantum walks, and this needs to be generalized to general 
quantum walks, or else find a counter example. 
One possible candidate is the graph 
consisting of two complete graphs connected by one edge.  
It is not clear that quantumly one cannot achieve convergence in time 
$O(n)$ which matches the $1/d\Phi=1/n\Phi$ lower bound. 

The limiting distribution 
for general quantum walks still needs to be understood. 
For Abelian groups, we have shown that coined quantum walks converge 
to the uniform distribution. 
On the other hand, we know one example
 in which a quantum walk does not converge 
to the same limiting distribution as the classical simple random walk. 
This is a quantum walk on the Cayley graph 
of the symmetric group $S_3$.
Is there a simple description, perhaps via representation 
theory, of the limiting distribution for 
quantum walks on Cayley graphs of non-Abelian groups? 

A very interesting question is how to use quantum walks
in order to speed up algorithms. 
One way to do that is via speeding up the convergence time,
however it is still an open question to give an example in which 
fast sampling cannot be done in an easy way classically.  
Another direction to pursue is to find other ways of
 using the various curious features 
of quantum walks, rather than speeding up the convergence time. 
For example, one might try to use quantum walks which converge 
to limiting distributions which are different than those of the
 corresponding classical walks.  Another way might be to investigate 
which quantum states can be generated using quantum walks. 
Generating interesting quantum states is an important primitive 
for quantum algorithms. A well known example is the graph 
isomorphism problem which can be reduced to the problem 
of generating a certain quantum state efficiently.

\section{Acknowledgements}
We wish to thank John Watrous for introducing us 
to his model of a quantum walk on a line. 
We are grateful to Barbara Terhal for useful discussions. 
We are most grateful to Alesha Kitaev for pointing out to us
 an error in a previous version of this paper.

\end{document}